\documentclass[%
 reprint,
 amsmath,amssymb,
 aps,
]{revtex4-2}

\usepackage{hyperref}
\usepackage{placeins}
\usepackage{graphicx}
\usepackage{dcolumn}
\usepackage{bm}
\usepackage{subcaption}

\begin{document}

\preprint{APS/123-QED}

\title{ Physics-informed neural network (PINN) modeling of charged particle multiplicity \\ using the two-component framework in heavy-ion collisions: A comparison with data-driven neural networks \\}

\author{Akash Das$^{1}$}
 \email{24pnpo01@iiitdmj.ac.in}
\author{Satya Ranjan Nayak$^{2}$}
 \email{satyanayak@bhu.ac.in}
\author{B. K. Singh$^{1,2}$}%
 \email{bksingh@bhu.ac.in}
 \email{director@iiitdmj.ac.in}
\affiliation{$^{1}$Discipline of Natural Sciences, PDPM Indian Institute of Information Technology Design \& Manufacturing, Jabalpur 482005, INDIA.\\
$^{2}$Department of Physics, Institute of Science,\\ Banaras Hindu University (BHU), Varanasi 221005, INDIA. }
\date{\today}

\begin{abstract}

In this study, we employ a conventional deep neural network (NN) framework integrated with physics-based constraints to predict charged hadron multiplicity ($N_{\text{ch}}$) in heavy-ion collisions. The goal is to assess the performance of a purely data-driven deep neural network in comparison to a physics-informed neural network (PINN). To accomplish this, we have taken data generated from the HYDJET++ model for testing and training purposes. We train our neural network frameworks using the data of one million individual $^{96}_{40}\text{Zr}+^{96}_{40}\text{Zr}$ collision events. Our PINN successfully extracts the hard-scattering fraction ($x$) by learning its underlying relation from the event data. For further testing and comparison with the conventional NN, we take data of $^{96}_{44}\text{Ru}+^{96}_{44}\text{Ru}$ (isobar of Zr) and $^{197}_{79}\text{Au}+^{197}_{79}\text{Au}$ collisions using the same simulation model. Once trained, the PINN demonstrates improved predictive performance on data not encountered during training, such as Au+Au collision results. Especially in regions of sparse data, corresponding to high $N_{\text{ch}}$ in our study, the PINN shows a clear advantage over a purely data-driven neural network.

\end{abstract}

\maketitle


\section{Introduction}

Heavy-ion collisions at relativistic energies provide a unique environment to study the properties of the quark-gluon plasma (QGP) \cite{Collins:1974ky, Shuryak:1978ij}. Experiments have been conducted on various colliding systems, such as $^{197}_{79}\text{Au}+^{197}_{79}\text{Au}$, $^{96}_{44}\text{Ru}+^{96}_{44}\text{Ru}$, and $^{96}_{40}\text{Zr}+^{96}_{40}\text{Zr}$
, to explore the collective behavior, nuclear structure properties, and isobaric effects \cite{STAR:2020xiv, STAR:2021mii}. The charged particle multiplicity ($N_{\text{ch}}$) and its correlations serve as a probe to know the initial geometry and particle production mechanism.

Relativistic Heavy Ion Collider (RHIC) and Large Hadron Collider (LHC) record particle tracks in detectors following high-energy nuclear collisions. These tracks are then correlated with the collision geometry using the Glauber model calculations \cite{PHENIX:2000owy, STAR:2001eyo}. The Glauber framework provides the number of nucleons taking part in the collision ($N_{\text{part}}$) and the number of binary nucleon-nucleon collisions ($N_{\text{coll}}$) corresponding to the impact parameter ($b$). The $N_{\text{ch}}$ is empirically related to $N_{\text{part}}$ and $N_{\text{coll}}$ via a two-component formula:

\begin{equation}
\label{eq:Nch}
\frac{dN_{\text{ch}}}{d\eta} = n_{pp} \left[ (1 - x) \cdot \frac{N_{\text{part}}}{2} + x \cdot N_{\text{coll}} \right]
\end{equation}

Here, $x$ represents the weight of the hard scattering contribution, and $n_{pp}$ is the average number of charged hadrons produced in individual proton-proton collisions. Since $N_{\text{part}}$ counts the total number of participating nucleons from both nuclei, the factor $1/2$ is introduced as a normalization convention so that the expression reduces to $n_{pp}$ in the proton-proton collision limit. The contributions from hard processes scale with $N_{\text{coll}}$ and soft processes with $N_{\text{part}}$.

In recent years, neural networks emerged as a promising tool in heavy-ion collision physics, enabling data-driven modeling of complex nonlinear dependencies \cite{Guest:2018yhq, Duarte:2020ngm}. Although effective, the standard neural networks (NN) rely purely on the statistical correlations of the data and lack physical interoperability. To overcome this limitation of conventional NN, a more advanced version of NN was introduced, known as Physics-Informed Neural Networks (PINN) \cite{Raissi2019}. This embeds physical laws directly into the NN's loss function and allows simultaneous learning from the data and theoretical constraints. 

In this work, we employed both a conventional data-driven neural network (NN) and a physics-informed neural network (PINN) to predict $N_{\text{ch}}$ in isobaric collisions. The neural networks are trained on Zr+Zr events generated from the HYDJET++ model \cite{Lokhtin:2008xi,lokhtin2010hydjetpp}, while testing is performed on Ru+Ru and Au+Au data generated within the same framework. As an isobar system, Ru+Ru collision data serves as a seen test dataset, in contrast the Au+Au collision data represent a completely unseen dataset during the training phase. The comparative analysis of NN and PINN predictions in these testing systems highlights the role of physics constraints in improving generalization across both seen and unseen collision systems. Additionally, we demonstrate that the PINN achieves better predictive performance than the purely data-driven NN even when trained on limited data, suggesting that physics constraints act as an effective regularizer when training samples are scarce. The study provides an initial step toward integrating PINN for analysis and prediction of observables in relativistic heavy-ion collisions. 

Accordingly, the objective of this work is to examine whether incorporating minimal, well-established physics constraints during training, such as the Glauber two-component formula in Eq.~(\ref{eq:Nch}), can improve the stability and generalization of neural-network models for event-by-event observables, particularly in sparse-data regimes.

For clarity, the purely data-driven neural network is referred to as NN (or normal NN), while the physics-informed neural network is denoted as PINN (or a neural network with physics imposed) throughout the rest of the paper. When referring to both NN and PINN collectively, we use the term “neural networks".

The article is organized as follows. In section-$\mathrm{II}$, we have discussed broadly about the NN architecture, Pythia, and HYDJET++ models. In section-$\mathrm{III}$, we provide the result and discussion part. In section-$\mathrm{IV}$, we have summarized our work.

\begin{figure*}[t]
  \centering
  \begin{minipage}{0.48\textwidth}
    \centering
    \includegraphics[width=\linewidth]{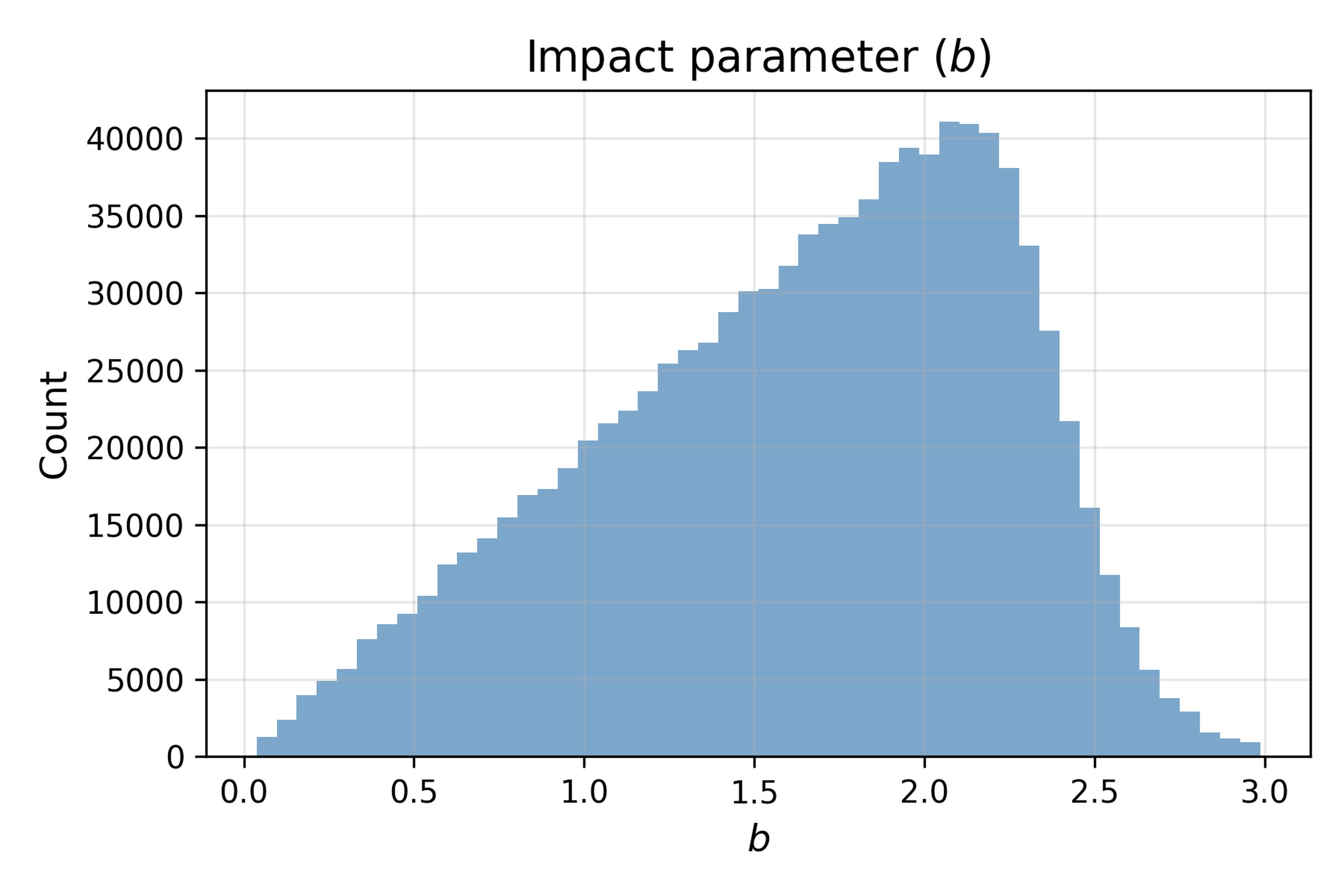}
    \subcaption{} \label{fig:b}
    \vspace{1em}
    \includegraphics[width=\linewidth]{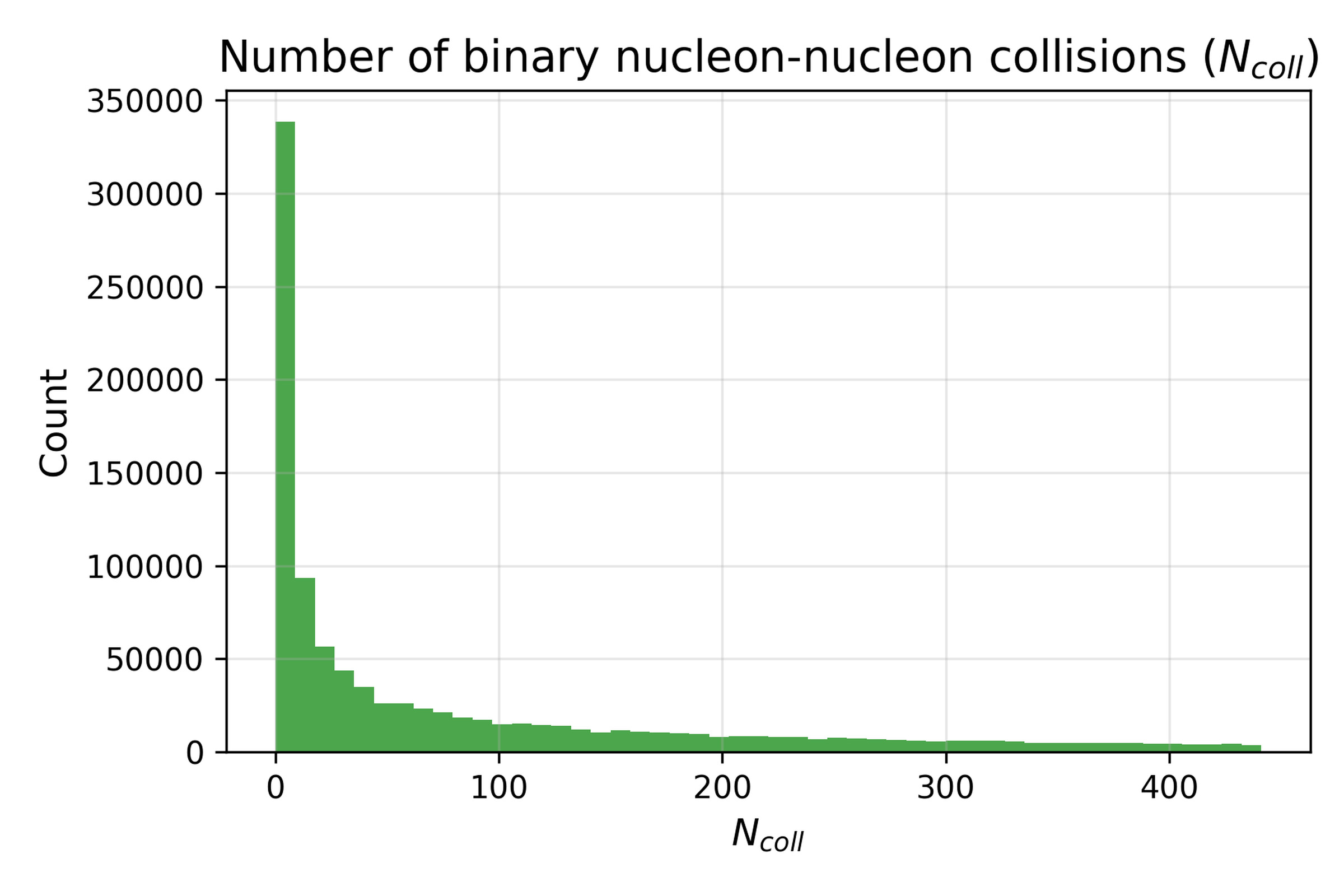}
    \subcaption{} \label{fig:ncoll}
  \end{minipage}\hfill
  \begin{minipage}{0.48\textwidth}
    \centering
    \includegraphics[width=\linewidth]{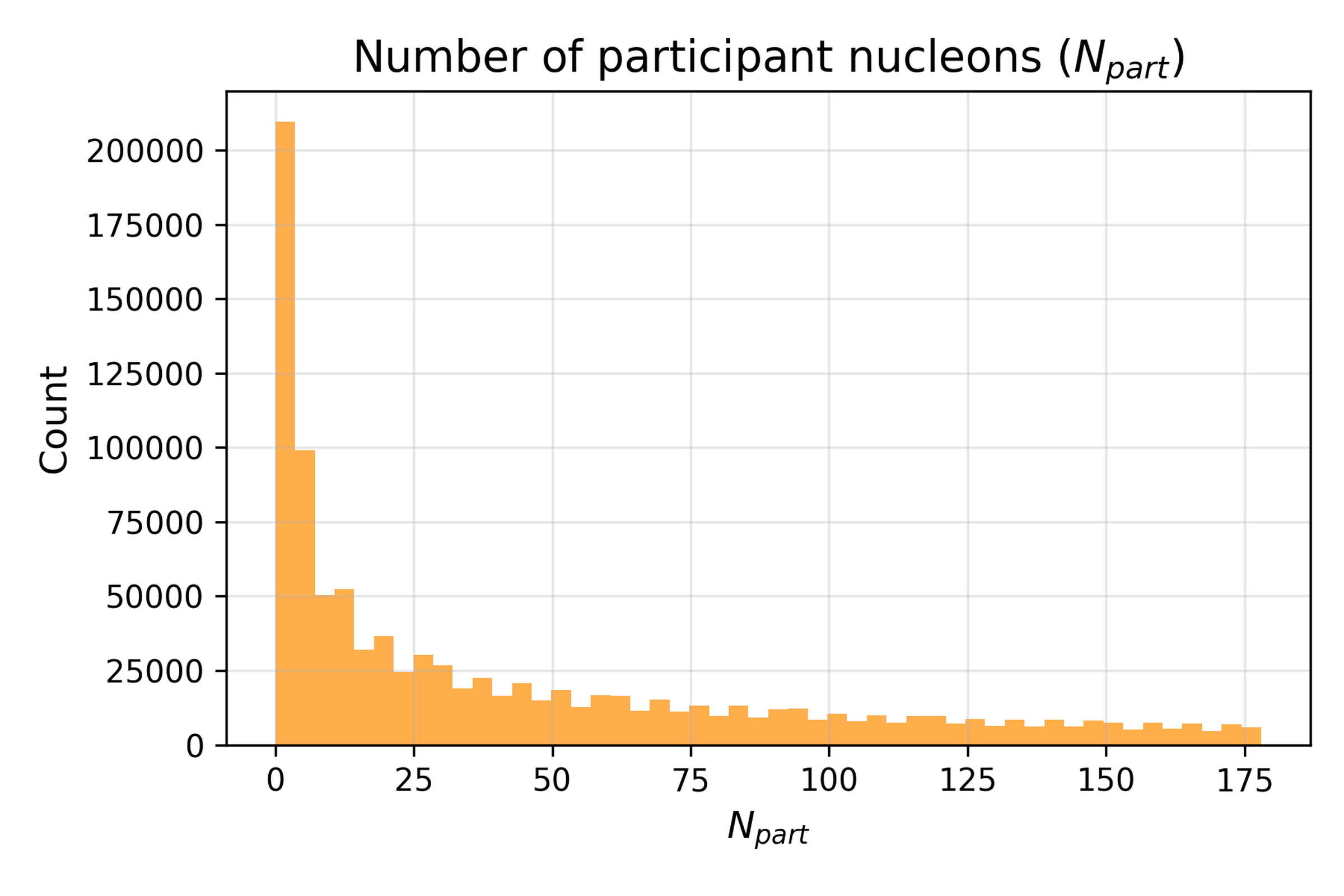}
    \subcaption{} \label{fig:npart}
    \vspace{1em}
    \includegraphics[width=\linewidth]{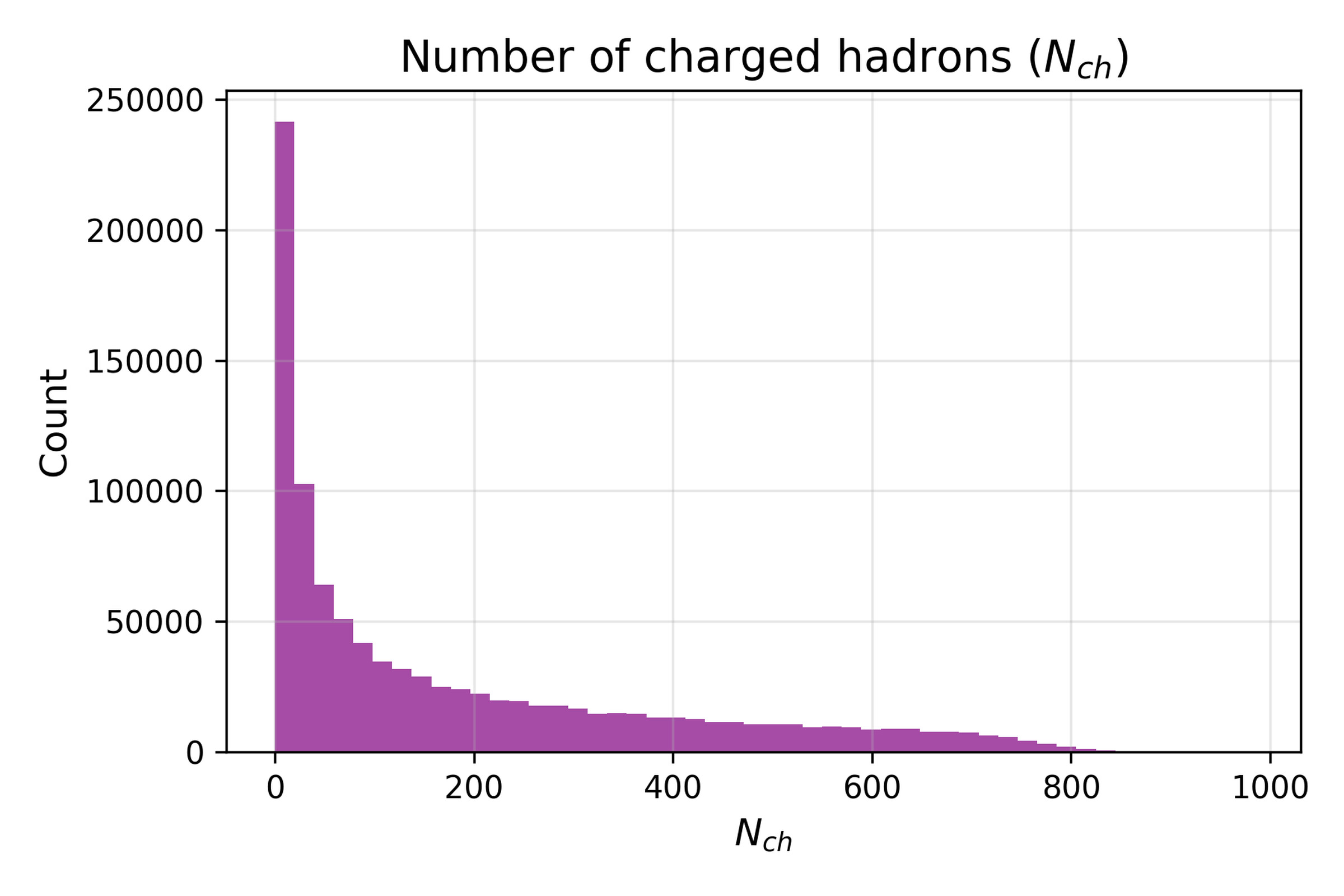}
    \subcaption{} \label{fig:nch}
  \end{minipage}
  
  \caption{Distributions of (\textbf{a}) impact parameter ($b$), (\textbf{b}) number of binary nucleon-nucleon collisions ($N_{coll}$), (\textbf{c}) number of participant nucleons ($N_{part}$), and (\textbf{d}) number of charged hadrons ($N_{ch}$) in the dataset.}
  \label{fig:data_distributions}
\end{figure*}

\section{Model Formalism}

\subsection{Physics framework}

The heavy-ion dataset was generated using the HYDJET++ event generator. HYDJET++ is a combination of two independent algorithms, i.e, PYQUEN for hard interactions and FASTMC for thermal particle production. The simulation starts by calculating the number of participants and the number of binary collisions using the Glauber model. Then the initial parton spectra are generated using PYTHIA6 \cite{Sjostrand:2006za}. PYQUEN incorporates the collisional and radiational energy loss of the partons inside the dense QGP medium. The thermal state is generated on freezeout hypersurfaces with preset freezeout conditions. The model assumes both chemical freezeout and thermal freezeout with $T_{ch}>T_{th}$ \cite{Amelin:2006qe,Amelin:2007ic}. The final hadronization is performed according to the LUND string fragmentation model \cite{Andersson:1997xwk,Andersson:2001yu}.

It is important to note that HYDJET++ doesn't include a fully dynamical hydrodynamic evolution. In this present work, HYDJET++ is employed as a controlled environment to generate large event samples. This allows us to isolate and study the role of physics-informed constraints in neural network training rather than to provide a complete physical description of the collision dynamics.

The multiplicity in pp collisions ($n_{pp}$) was generated using PYTHIA 8 \cite{Bierlich:2018xfw,bierlich2022scipost}. PYTHIA is an all-purpose Monte Carlo event generator used to simulate a wide range of collision systems. The hard interaction between the valency quarks is calculated using standard pQCD tools. The softer interactions between the beam remnants are simulated using the Multi-Parton interaction (MPI) framework. A detailed description can be found in the corresponding papers \cite{helenius2019epjc}. The parton showers are simulated simultaneously and added back to the final state. A color reconnection \cite{Sjostrand:1987su} is performed between the $q\bar{q}$ dipoles to reduce the number of long strings and control the multiplicity. Finally, the strings are fragmented into hadrons according to the Lund string fragmentation model \cite{Andersson:1997xwk,Andersson:2001yu}. A detailed description of PYTHIA can be found in the PYTHIA manual \cite{Sjostrand:2006za}.

\subsection{Model Architecture and Data Processing}

In this study, we employed the PyTorch deep learning framework \cite{paszke2019pytorch} to construct and train the PINN for predicting $N_{\text{ch}}$ based on event-level observables. Here, event-level observables denote the $N_{\text{part}}$, $N_{\text{coll}}$ and the impact parameter ($b$) for each collision event. PyTorch’s ecosystem offers a flexible and modular platform for implementing the NN framework. PyTorch neural network architectures are typically implemented by sub-classing \texttt{torch.nn.Module}, the base class that organizes network layers and automatically registers trainable parameters for gradient-based optimization. PyTorch further provides core features such as GPU acceleration, automatic differentiation, learning rate scheduling, and gradient clipping etc.

For training the neural networks, we employ a dataset generated from simulated Zr+Zr collision events at RHIC energies using the HYDJET++ event generator. The dataset consists of four features per event:- $b$, $N_{\text{part}}$, $N_{\text{coll}}$, and $N_{\text{ch}}$. The input vector X = (b, $N_{\text{part}}$, $N_{\text{coll}}$) is fed into a fully connected feedforward network, while the target variable Y = $N_{\text{ch}}$ serves as a prediction target. 

Before training, all the input features and the target variables are normalized to the range $[0,1]$ using min-max scaling \cite{Patro:2015Normalization}. The normalized data is converted to PyTorch tensors and split into an $80$\%-$20$\% train-test partition using \texttt{random\_split}. The batch size is adjusted according to the size of the training dataset. For a training dataset of one million samples, a batch size of 1000 is used to ensure efficient training using PyTorch's \texttt{DataLoader} \cite{paszke2019pytorch}.

The neural network architecture consists of an input layer of dimension 3, followed by three hidden layers containing 128, 64, and 32 neurons, respectively. The first hidden layer uses a hyperbolic tangent (\texttt{tanh}) activation function, followed by  Rectified Linear Unit (\texttt{ReLU}) activations in subsequent layers \cite{Dubey:2021activation}. Dropout layers with a rate of 0.1 are applied after the second and third hidden layers to mitigate overfitting. A learning rate of 0.0001 is employed for training, and the mean squared error (MSE) is used as the loss function. The output layer produces a single scalar prediction corresponding to $N_{\text{ch}}$.

A physics module is incorporated in parallel with the data-driven neural network to enforce known physical constraints. This module implements a fixed analytical expression for charged hadron multiplicity inspired by the Glauber two-component formula Eq.~(\ref{eq:Nch}). In this module, $n_{\text{pp}}$ is treated as a non-trainable buffer using \texttt{register\_buffer}. On the other hand, $x$ is taken as a scaler trainable parameter defined as \texttt{self.x = nn.Parameter(torch.tensor(1.0, dtype=torch.float32))}. The value $1.0$ is chosen as a neutral initial value within the physically expected range $[0,1]$. Since it is wrapped with \texttt{nn.Parameter}, it is registered as a model parameter and included in the gradient-based optimization.

The overall training is guided by a composite loss function:

\begin{equation}
\mathcal{L}_{\text{total}} = \mathcal{L}_{\text{data}} + \lambda \, \mathcal{L}_{\text{physics}}
\end{equation}

The $\mathcal{L}_{\text{data}}$ term quantifies the deviation between the neural-network–predicted charged-particle multiplicity and the simulated event-by-event data. The $\mathcal{L}_{\text{physics}}$ term penalizes deviations of the data-driven network prediction from the charged-particle multiplicity computed using the Glauber two-component relation (Eq.~(\ref{eq:Nch})) for the same event-level inputs. Both losses use the \texttt{MSELoss} function to ensure robustness against the outliers. The hyperparameter $\lambda$ adjusts how much importance we give to the physics rules in matching the data. 

The training was carried out on Google Colab using an NVIDIA T4 accelerator with 16 GB of memory. The T4 provides hardware acceleration optimized for deep learning, enabling efficient parallel computations. PyTorch was employed as the deep learning framework, leveraging Colab’s GPU/TPU backend for faster neural network training and experimentation.

\begin{figure}[htbp]
  \centering
  \includegraphics[width=0.9\linewidth]{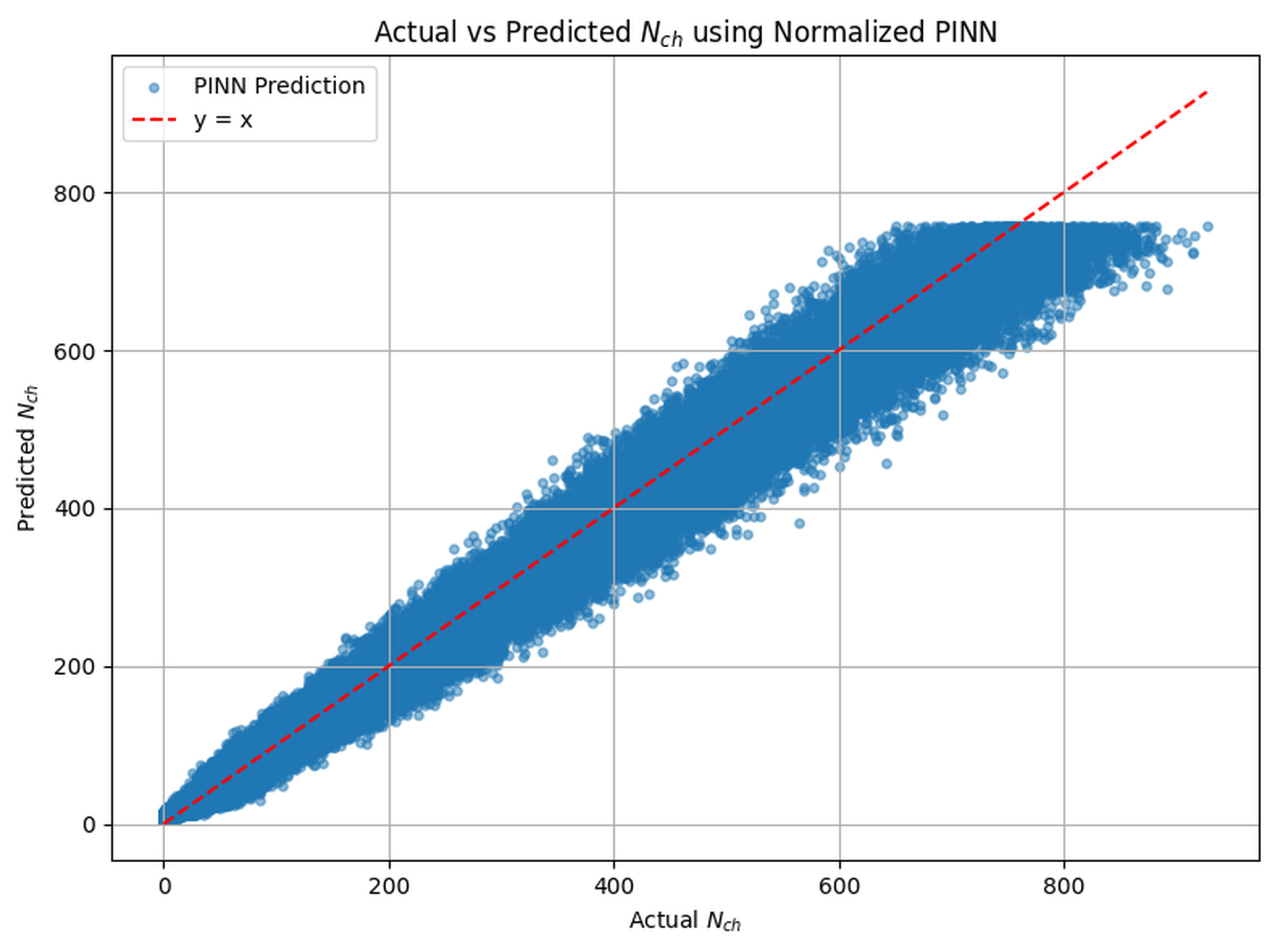} 
 \caption{Comparison of simulated (actual) and PINN-predicted $N_{\mathrm{ch}}$ for Zr+Zr, obtained with $\lambda = 0.1$ and $x = 0.41$.}
\end{figure}

\begin{figure*}[t]
  \centering
  \begin{minipage}{0.48\textwidth}
    \centering
    \includegraphics[width=\linewidth]{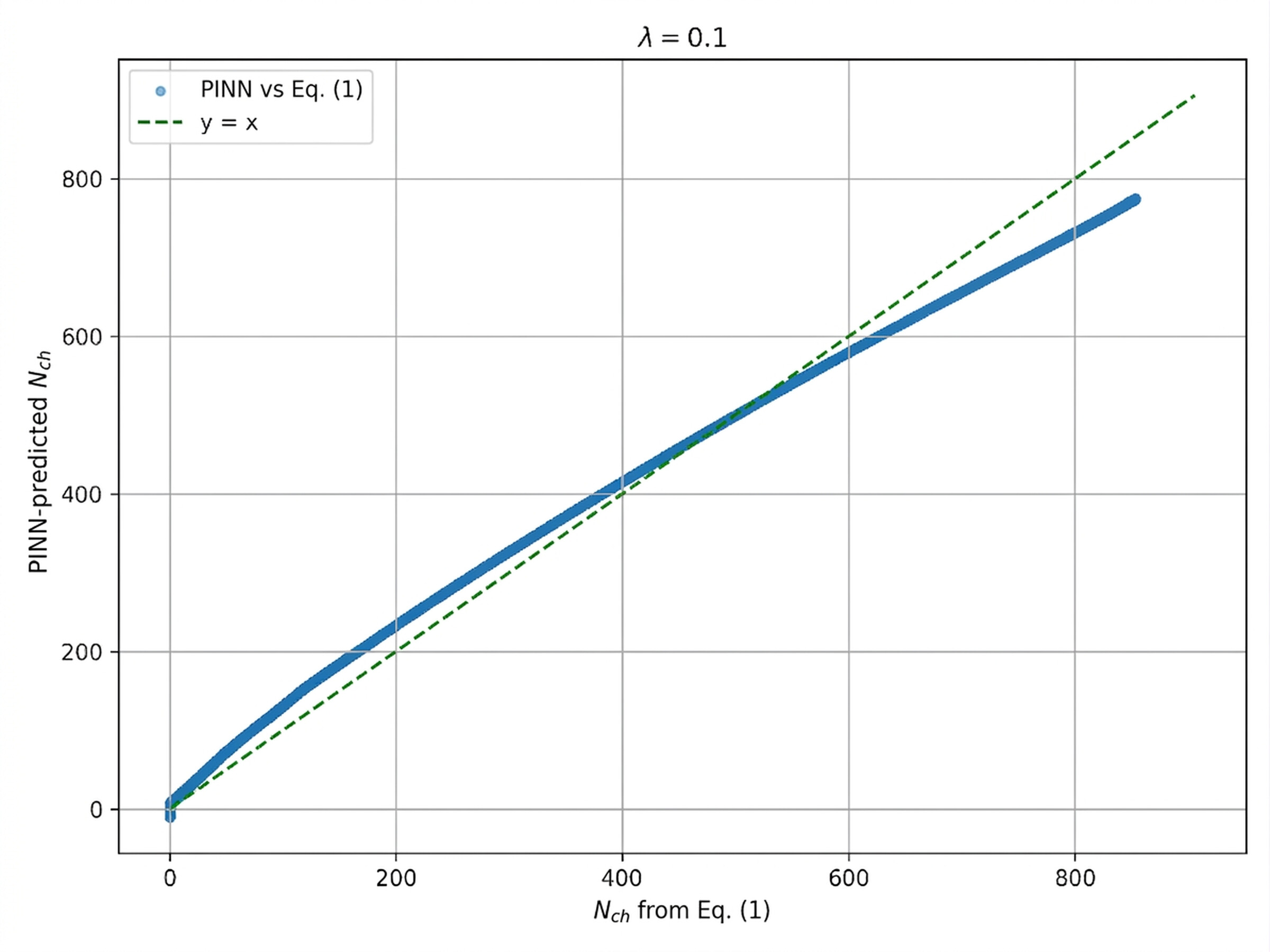}
    \subcaption{} \label{fig:a}
    \vspace{1em}
  \end{minipage}\hfill
  \begin{minipage}{0.48\textwidth}
    \centering
    \includegraphics[width=\linewidth]{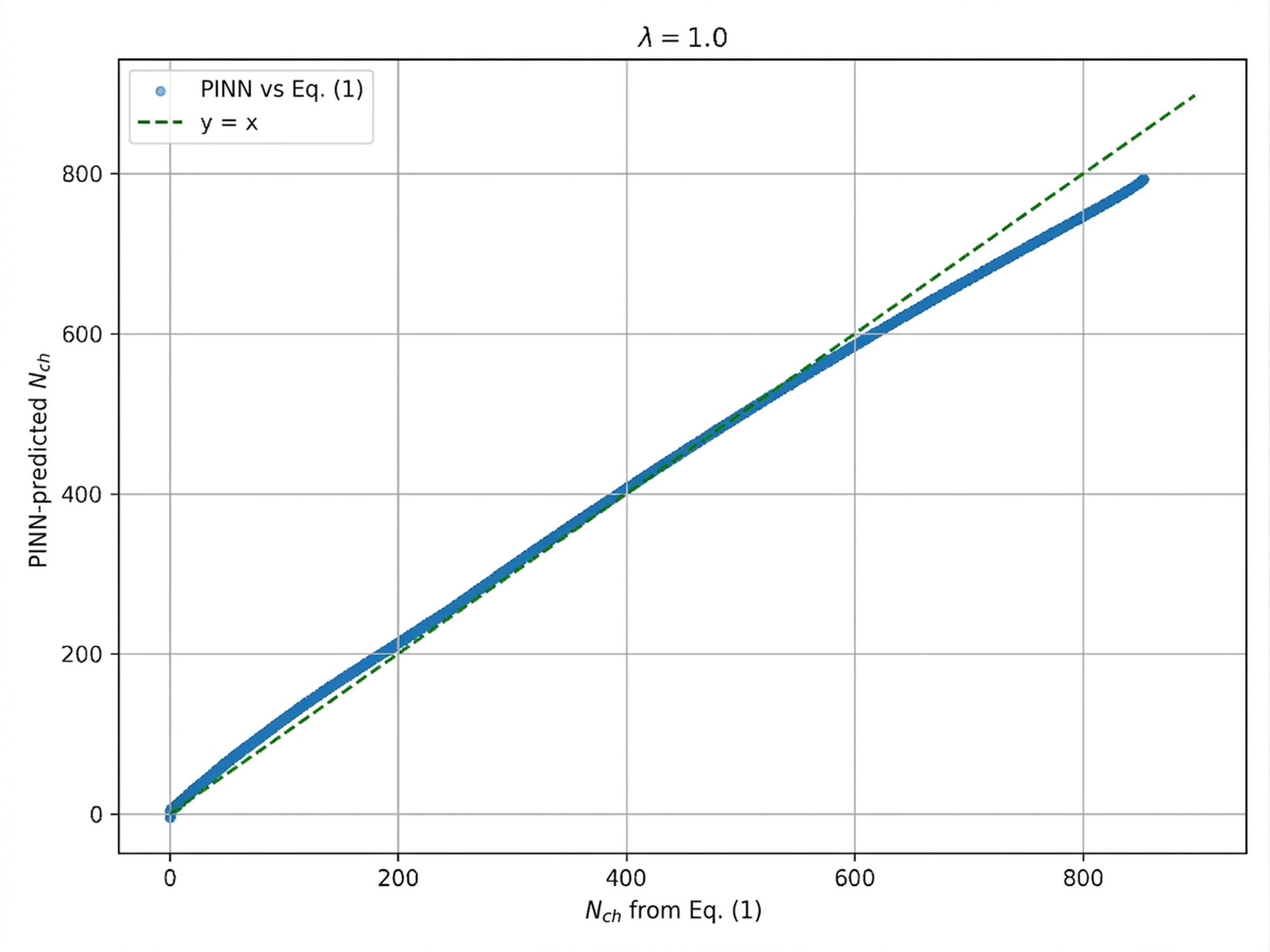}
    \subcaption{} \label{fig:c}
    \vspace{1em}
  \end{minipage}
  
  \caption{Comparison of PINN-predicted $N_{\text{ch}}$ and $N_{\text{ch}}$ calculated from Eq. (1) for (a) $\lambda$ = 0.1 and (b) $\lambda$ = 1.0 for Zr+Zr datasets.}
  \label{fig:lambda_comparison}
\end{figure*}

\begin{table}[htbp]
\centering

\label{tab:performance}
\small 
\begin{tabular}{lcc}
\hline
\textbf{Metric} & \(\boldsymbol{\lambda = 0.1}\) & \(\boldsymbol{\lambda = 1}\) \\
\hline
\(R^{2}\)  & 0.9880 & 0.9841 \\
RMSE    & 23.16 & 26.07 \\
MAE    & 14.85 & 17.16 \\
\hline
\end{tabular}
\caption{Performance metrics for different values of $\lambda$ with $x = 0.41$ and $n_{pp} = 3.602$, trained and tested on $0.1$ million Zr+Zr data points for $100$ epochs.}
\end{table}

\section{Result and discussions}

The input data distributions are shown in Fig. 1. An intrinsic non-uniformity is present in the data. In the Glauber model, events are sampled with a probability proportional to \( 2\pi b \, db \), so smaller \( b \) values are less probable than intermediate ones. The distribution peaks around $b \sim 2$ fm before dropping off due to the finite nuclear size. Since $N_{\text{part}}$, $N_{\text{coll}}$, and $N_{\text{ch}}$ are all governed by the collision geometry and centrality, their distributions exhibit similar centrality-dependent trends. Due to this, our neural-networks has fewer training samples in the high-$N_{\text{ch}}$ region. To handle this sparse-data regime, a PINN is a good choice.

A total of $10^6$ minimum bias Zr+Zr collision events were simulated at $\sqrt{s_{NN}}=200$ GeV using the HYDJET++ model, assuming spherical nuclear configurations obtained from the density functional theory (DFT) calculations \cite{Xu:2021vpn}. Those events are selected that fall under the pseudorapidity cut $|\eta| < 0.5$. 

The code for purely data-driven and physics-based frameworks is nearly the same, with the latter incorporating an additional physics-informed loss term. We started with learning the value of $x$ using PINN. For that, we took $x$ as a trainable parameter in our code. The value of $n_{pp}$ is made fixed to 3.602 as obtained with the help of PYTHIA. The PINN trains on $1$ million minimum-bias Zr+Zr collision data points with $\lambda = 0.1$, converged within $32$ epochs (early stopped), yielding a predicted value of $x \approx 0.4097$, which we round off to $0.41$. As discussed earlier, the $\mathcal{L}_{\rm physics}$ penalizes the deviation between the Glauber two-component formula and the NN prediction. The NN weights are updated through both $\mathcal{L}_{\rm data}$ and $\mathcal{L}_{\rm physics}$, while $x$ is updated only through $\mathcal{L}_{\rm physics}$. Consequently, $x$ converges to the hard-scattering fraction that best reconciles the Glauber formula with the simulated event-level $N_{\mathrm{ch}}$ values learned by the NN. To validate this predicted value of $x$, $x$ (together with $n_{pp}$) was fixed at externally chosen values and scanned over the range $0.1 \leq x \leq 1.0$ to verify that $x = 0.41$ provides the best agreement with the data. The corresponding parity plot is shown in Fig. 2.

\begin{figure}[htbp]
  \centering
  \includegraphics[width=0.9\linewidth]{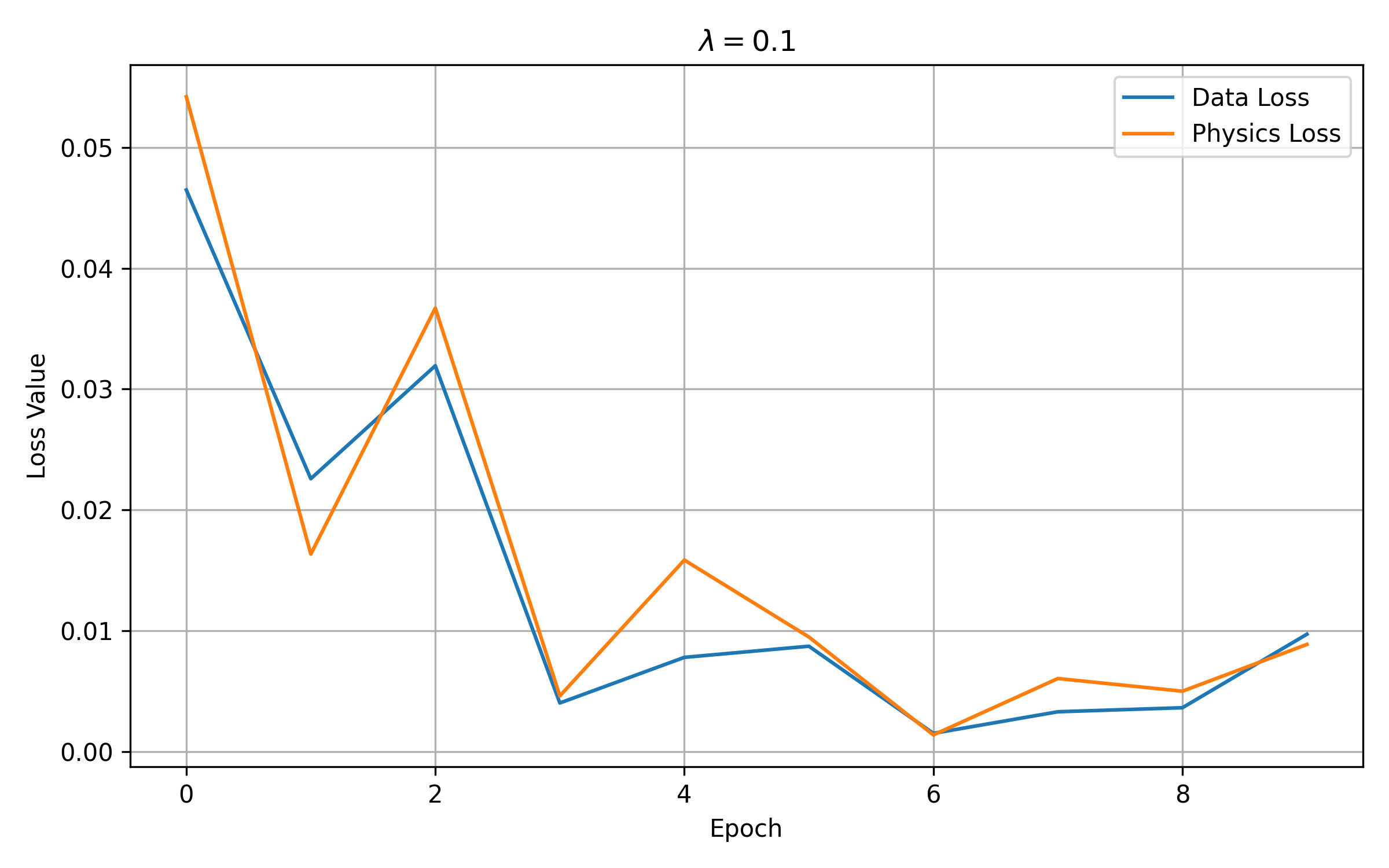} 
  \caption{Loss trends for 500 training data-sets of Zr+Zr collisions.}
\end{figure}

\begin{figure}[htbp]
  \centering
  \includegraphics[width=0.9\linewidth]{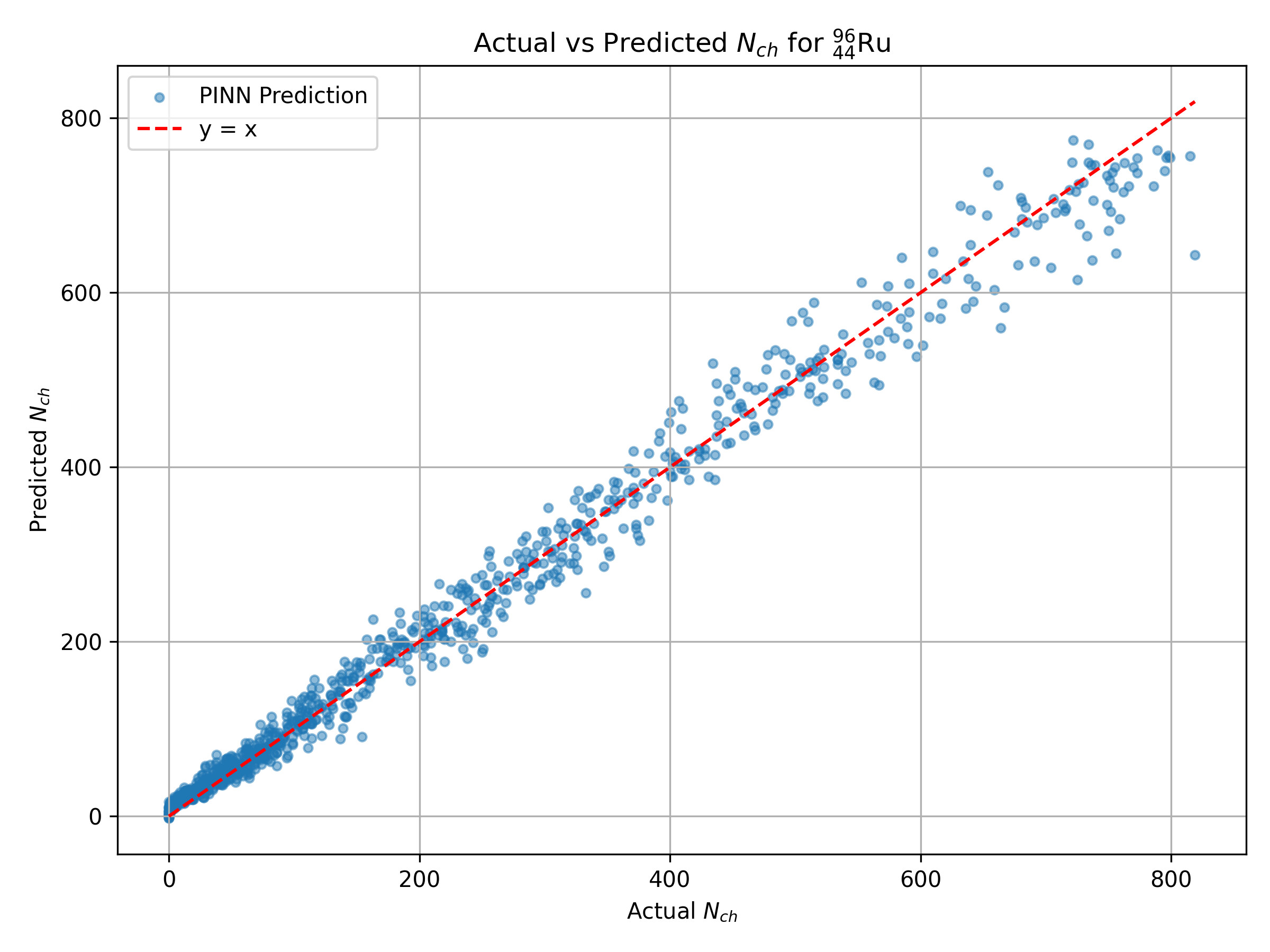} 
  \caption{Comparison of simulated (actual) and data-driven NN predicted $N_{\text{ch}}$ for the test data of Ru+Ru.}
\end{figure}

\begin{figure}[htbp]
  \centering
  \includegraphics[width=\linewidth]{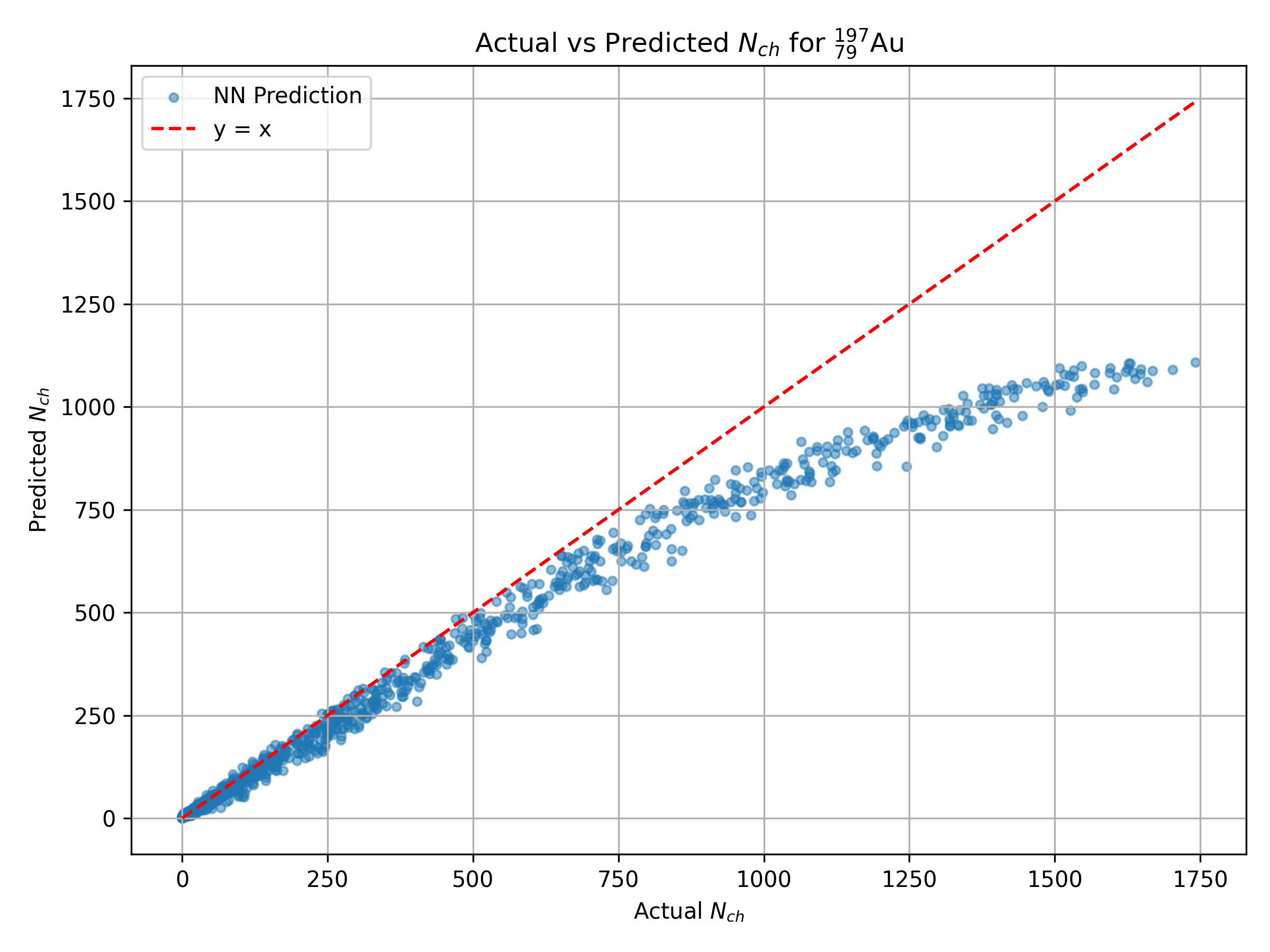}
  \centerline{(a)}\vspace{1ex}
  \includegraphics[width=\linewidth]{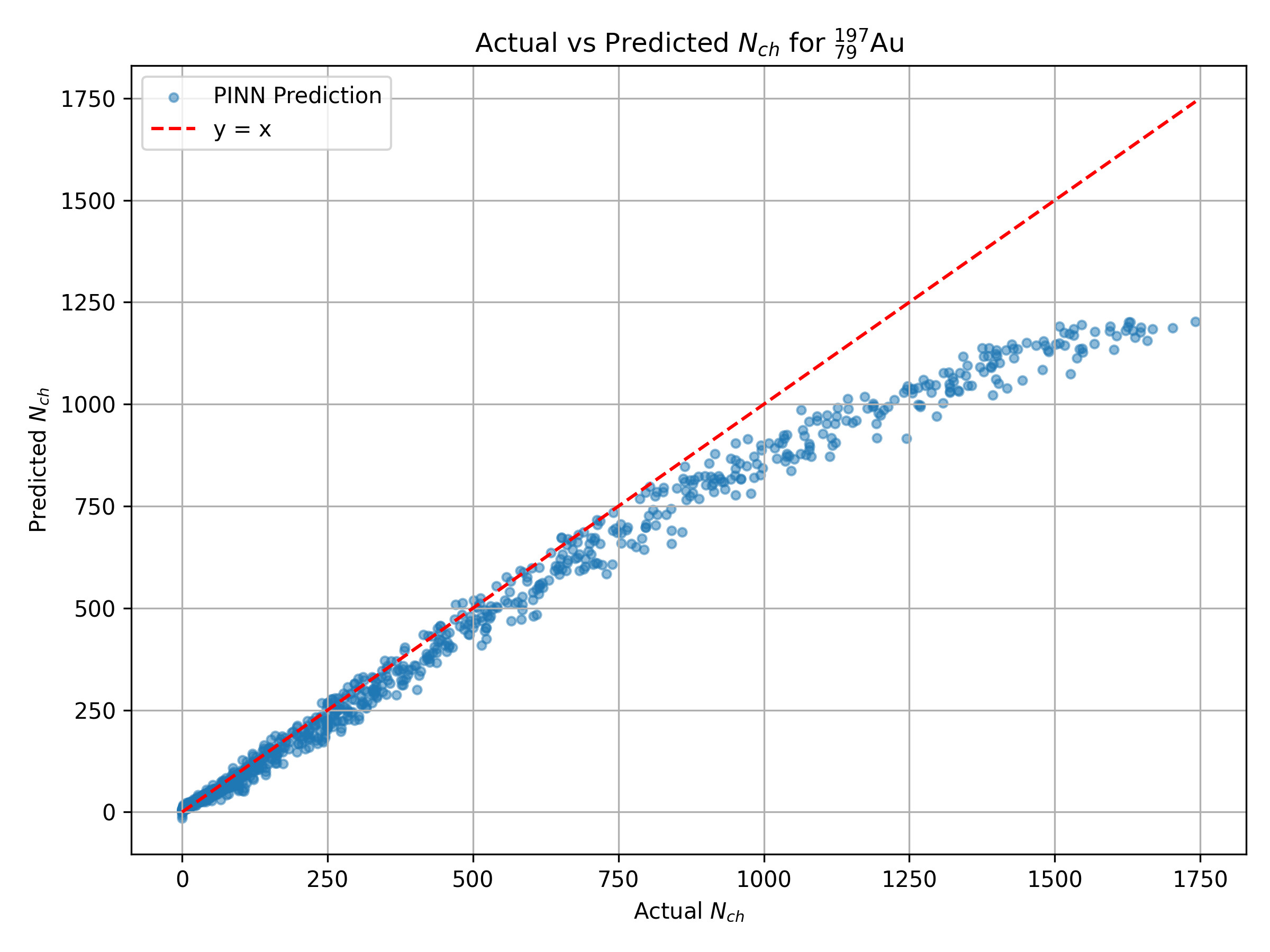}
  \centerline{(b)}\vspace{1ex}
  \includegraphics[width=\linewidth]{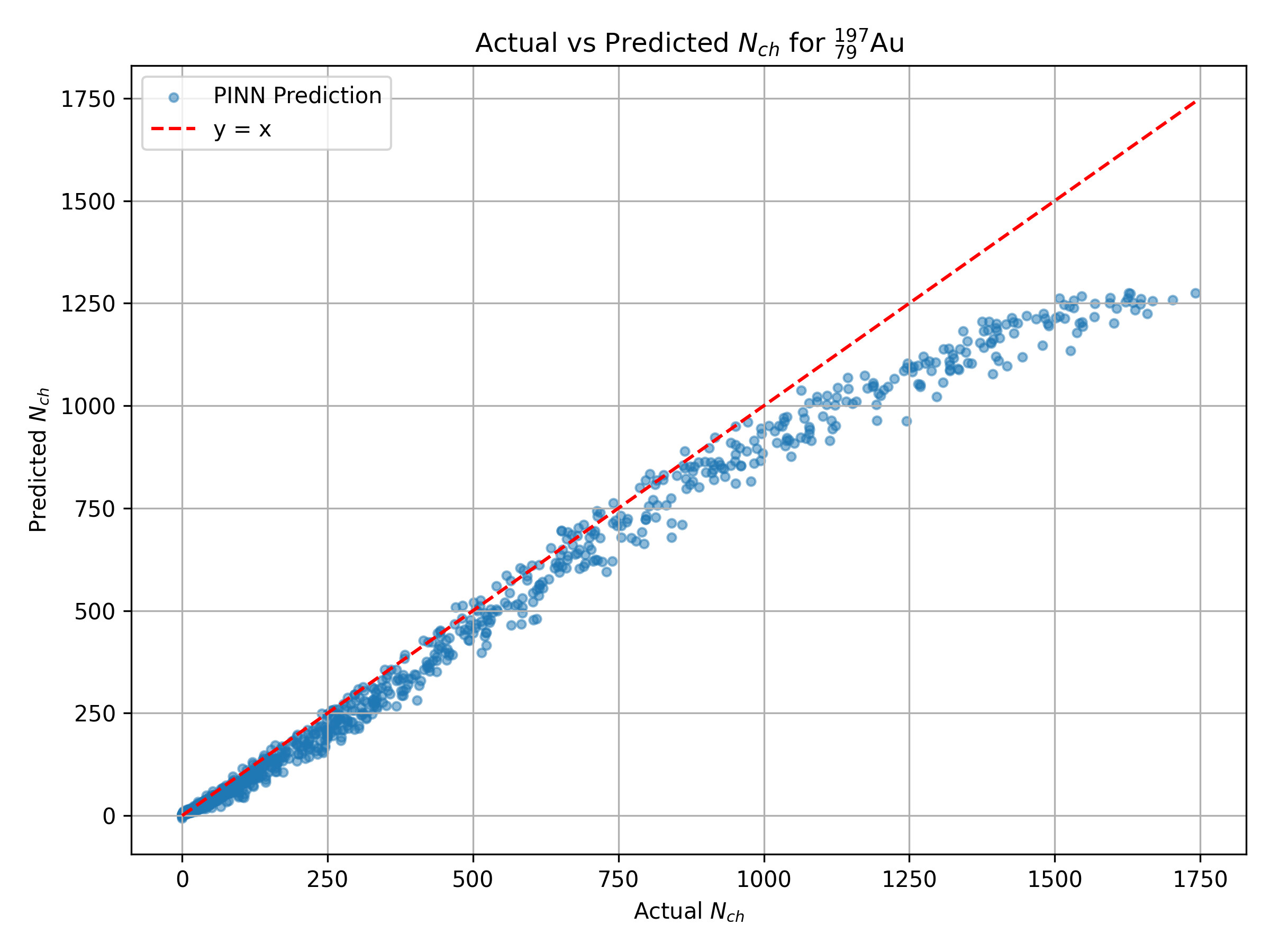}
  \centerline{(c)}
  
  \caption{Comparison of simulated (actual) and model-predicted $N_{\text{ch}}$ for the Au+Au test data: (a) without physics, (b) $\lambda=0.1$, and (c) $\lambda=1.0$.}
  \label{fig:Au_comparison}
\end{figure}

Taking $x = 0.41$, we conducted trials to determine the value of $\lambda$ that provides a good balance between data loss and physics loss (Table 1). The corresponding plots are presented in Fig. 3, obtained using one-tenth of the original dataset (0.1 million events) of Zr+Zr collisions and trained for 100 epochs. In panels (a) and (b), the x-axis shows $N_{\mathrm{ch}}$ computed directly from the Glauber two-component formula (Eq. 1), while the y-axis shows the PINN-predicted $N_{\mathrm{ch}}$ for the same event-level inputs. As $\lambda$ increases, the composite loss function increasingly penalizes deviation from the physics formula, causing the PINN output to align more closely with the $N_{\mathrm{ch}}$ calculated from Eq.~(1). The near-linear relationship in mid-central to peripheral collisions ($0 \leq N_{\mathrm{ch}} \leq 400$) seen in panel (b) compared to panel (a) confirms this. This particular behavior is again evident in the high-$N_{\mathrm{ch}}$ region ($N_{\mathrm{ch}} \gtrsim 600$), where training data is sparse due to Glauber geometry statistics. In this region, the data loss provides weak gradient guidance, hence the PINN relies increasingly on the physics constraint to maintain physically consistent predictions. Thus the physics constraint is being correctly enforced, especially where data alone is insufficient.

The comparison between PINN and purely data-driven NNs is performed using limited Zr+Zr data to evaluate how much data is sufficient for effective training. Corresponding evaluation matrices are tabulated in Table~II. For training and testing, only 500 data points are used on an 80–20\% train-test split. A batch size of 10 is employed and the value of $\lambda$ is taken as $0.1$. We tested the neural-network with and without physics for 10 epochs. Table II shows a consistent improvement across all three metrics when physics constraints are incorporated. The loss curve for this is plotted in Fig. 4, where the physics loss remains above the data loss throughout training. Notably, even this partial enforcement of the physics-based loss is sufficient to yield measurably better predictions than the purely data-driven network. This suggest that, the physics term provides a useful regularization effect in the limited-data regime.

\begin{table}[!ht]
\centering
\label{tab:performance_500_vertical_split}
\small
\begin{tabular}{l|c|c} 
\hline
\multicolumn{3}{c}{\textbf{10 Epochs}} \\
\hline
\textbf{} & \textbf{PINN} & \textbf{Normal NN} \\
\hline
$R^2$ & 0.9789 & 0.9742 \\
RMSE & 31.19 & 34.68 \\
MAE & 20.30 & 23.75 \\
\hline
\end{tabular}
\caption{Performance metrics of PINN and normal NN trained on 500 data points over 10 epochs.}
\end{table}

The neural-networks trained on Zr+Zr collisions are evaluated on Ru+Ru and Au+Au data to assess generalizability. We trained separate PINNs, with physics weights $\lambda = 0.1$ and $\lambda = 1.0$, as well as a purely data-driven neural network, all on one million Zr+Zr events over 100 epochs. Testing datasets of 1000 events each are taken from Ru+Ru and Au+Au collisions. The purely data-driven neural network, without physics-based constraint, predicts the Ru's $N_{\text{ch}}$ with high accuracy, as it is an isobar of Zr (as shown in Fig. 5). We further evaluated the trained networks on Au data, which was completely absent during the training process, serving as a fully unseen test set. In Fig. 6, the predicted $N_{\text{ch}}$ of Au is shown for three different conditions, namely (a) without physics constraints, (b) with PINN ($\lambda = 0.1$), and (c) with PINN ($\lambda = 1.0$). Without physics-based loss, the neural-network starts underpredicting around $N_{\mathrm{ch}} \approx 500$. This indicates that the purely data-driven NN is poor at unknown systems. Meanwhile, with physics imposed in training, the model predicts better for Au+Au data. The prediction accuracy improves for $\lambda = 1.0$ compared to $\lambda = 0.1$, as the former predicts higher $N_{\text{ch}}$ values (till $N_{\mathrm{ch}} \approx 1100$) that are not present in the Zr's training dataset. 

Nevertheless, the PINN does not fully reproduce the Au's most central multiplicity distribution. This is expected as the model was trained exclusively on Zr+Zr collision data where the maximum observed $N_{\mathrm{ch}}$ is approximately $800$. Whereas in central Au+Au collisions, the yield $N_{\mathrm{ch}}$ goes up to $\sim 1750$, with $N_{\mathrm{part}}$ and $N_{\mathrm{coll}}$ values approximately $2$--$3$ times larger than Zr. Furthermore, the value of $x$ is optimized for Zr+Zr collisions and is not expected to fully capture the scaling behavior of the heavier Au + Au collisions. Despite these, the PINN still significantly outperforms the purely data-driven neural network, suggesting that the model has learned aspects of the underlying physical constraints and can generalize this behavior across different collision systems. This demonstrates the point that even a minimal physics constraint can provide meaningful extrapolation capability beyond the training domain.

\section{Summary}

In summary, we started training our neural-networks with Zr+Zr collisions data generated by the HYDJET++ model. A limited number of data points are present in the low impact parameter region, due to Glauber statistics. So, the use of a physics-informed neural network (PINN) helps to compensate for this scarcity by incorporating physics constraints into the learning process. We have taken $x$ as a scalar trainable parameter, and the PINN predicts a value of $x$=0.41. We then verified this value of $x$ by performing a validation scan in which $x$ is not optimized, but instead fixed to discrete externally chosen values for each run. With the increasing value of $\lambda$, the PINN output shows stronger agreement with the Glauber two-component formula prediction, which reflects greater enforcement of the physics constraint. We also find that physical constraints can make the NN learn better with fewer training data points. For predicting the charged-particle multiplicity in isobaric Ru+Ru collisions, a purely data-driven neural network is sufficient. However, for extrapolating to collision systems not encountered during training, a physics-informed neural network is required. With the PINN, the trained model demonstrates improved generalization to unseen test datasets, as illustrated by the previously unseen Au+Au collisions data.

We emphasize that the present study is intended to demonstrate the advantages of PINN-based methodology over pure data-driven neural-networks, rather than to provide a final predictive description of heavy-ion collision observables. Thus, it is expected that our work will enlighten the role of PINN over conventional NN in heavy-ion collision problems. PINN can specifically be effective at reducing the number of simulated events and the overall computation time in heavy-ion collision phenomenological analysis. PINN can also help predict the outcomes of Beam Energy Scan (BES) studies, provided that data at specific energies and reliable theoretical constraints are available.

\section{Data availability}

Data and codes will be made available on request.

\section{Acknowledgment}

We are grateful to Dr. Stephen Baek for his valuable lectures on PINN and L. Pon Vijaya Kanthan for his guidance on coding. BKS sincerely acknowledges financial support from the Institute of Eminence (IoE), BHU Grant number 6031. AD acknowledges the financial support through the Institute fellowship from IIITDM Jabalpur. SRN acknowledges the financial support from the UGC
Non-NET fellowship and IoE research incentive during
the research work.


\begin{thebibliography}{46}

\bibitem{Collins:1974ky}
J.~C.~Collins and M.~J.~Perry,
Phys. Rev. Lett. \textbf{34} (1975), 1353
doi:10.1103/PhysRevLett.34.1353



\bibitem{Shuryak:1978ij}
E.~V.~Shuryak,
Phys. Lett. B \textbf{78} (1978), 150
doi:10.1016/0370-2693(78)90370-2




\bibitem{STAR:2021mii}
M.~Abdallah \textit{et al.} [STAR],
Phys. Rev. C \textbf{105} (2022) no.1, 014901
doi:10.1103/PhysRevC.105.014901
[arXiv:2109.00131 [nucl-ex]].




\bibitem{STAR:2020xiv}
J.~Adam \textit{et al.} [STAR],
Phys. Rev. C \textbf{102} (2020) no.5, 054913
doi:10.1103/PhysRevC.102.054913
[arXiv:2006.00582 [nucl-ex]].



\bibitem{PHENIX:2000owy}
K.~Adcox \textit{et al.} [PHENIX],
Phys. Rev. Lett. \textbf{86} (2001), 3500-3505
doi:10.1103/PhysRevLett.86.3500
[arXiv:nucl-ex/0012008 [nucl-ex]].



\bibitem{STAR:2001eyo}
C.~Adler \textit{et al.} [STAR],
Phys. Rev. Lett. \textbf{87} (2001), 112303
doi:10.1103/PhysRevLett.87.112303
[arXiv:nucl-ex/0106004 [nucl-ex]].



\bibitem{Guest:2018yhq}
D.~Guest, K.~Cranmer and D.~Whiteson,
Ann. Rev. Nucl. Part. Sci. \textbf{68} (2018), 161-181
doi:10.1146/annurev-nucl-101917-021019
[arXiv:1806.11484 [hep-ex]].



\bibitem{Duarte:2020ngm}
J.~Duarte and J.~R.~Vlimant,
doi:10.1142/9789811234033{\_}0012
[arXiv:2012.01249 [hep-ph]].




\bibitem{Raissi2019}
M.~Raissi, P.~Perdikaris and G.~E.~Karniadakis,
J.\ Comput.\ Phys.\ \textbf{378}, 686--707 (2019).
doi:10.1016/j.jcp.2018.10.045





\bibitem{Lokhtin:2008xi}
I.~P.~Lokhtin, L.~V.~Malinina, S.~V.~Petrushanko, A.~M.~Snigirev, I.~Arsene and K.~Tywoniuk,
Comput. Phys. Commun. \textbf{180} (2009), 779-799
doi:10.1016/j.cpc.2008.11.015
[arXiv:0809.2708 [hep-ph]].




\bibitem{lokhtin2010hydjetpp}
I.~P.~Lokhtin, L.~V.~Malinina, S.~V.~Petrushanko and A.~M.~Snigirev,
Phys.\ Atom.\ Nucl.\ \textbf{73} (2010), 2139--2147
doi:10.1134/S1063778810120203
\href{https://doi.org/10.1134/S1063778810120203}{[DOI]}




\bibitem{Sjostrand:2006za}
T.~Sjostrand, S.~Mrenna and P.~Z.~Skands,
JHEP \textbf{05} (2006), 026
doi:10.1088/1126-6708/2006/05/026
[arXiv:hep-ph/0603175 [hep-ph]].




\bibitem{Amelin:2006qe}
N.~S.~Amelin, R.~Lednicky, T.~A.~Pocheptsov, I.~P.~Lokhtin, L.~V.~Malinina, A.~M.~Snigirev, I.~A.~Karpenko and Y.~M.~Sinyukov,
Phys. Rev. C \textbf{74} (2006), 064901
doi:10.1103/PhysRevC.74.064901
[arXiv:nucl-th/0608057 [nucl-th]].



\bibitem{Amelin:2007ic}
N.~S.~Amelin, R.~Lednicky, I.~P.~Lokhtin, L.~V.~Malinina, A.~M.~Snigirev, I.~A.~Karpenko, Y.~M.~Sinyukov, I.~Arsene and L.~Bravina,
Phys. Rev. C \textbf{77} (2008), 014903
doi:10.1103/PhysRevC.77.014903
[arXiv:0711.0835 [hep-ph]].




\bibitem{Andersson:1997xwk}
B.~Andersson,
Camb. Monogr. Part. Phys. Nucl. Phys. Cosmol. \textbf{7} (1997), 1-471
Cambridge University Press, 1998,
ISBN 978-1-009-40129-6, 978-1-009-40125-8, 978-1-009-40128-9, 978-0-521-01734-3, 978-0-521-42094-5, 978-0-511-88149-7
doi:10.1017/9781009401296





\bibitem{Andersson:2001yu}
B.~Andersson, S.~Mohanty and F.~Soderberg,
Eur. Phys. J. C \textbf{21} (2001), 631-647
doi:10.1007/s100520100757
[arXiv:hep-ph/0106185 [hep-ph]].




\bibitem{Bierlich:2018xfw}
C.~Bierlich, G.~Gustafson, L.~L{\"o}nnblad and H.~Shah,
JHEP \textbf{10} (2018), 134
doi:10.1007/JHEP10(2018)134
[arXiv:1806.10820 [hep-ph]].



\bibitem{bierlich2022scipost}
C.~Bierlich, S.~Chakraborty, N.~Desai, L.~Gellersen, I.~Helenius, P.~Ilten, L.~Lönnblad,
S.~Mrenna, S.~Prestel, C.~T.~Preuss \textit{et al.},
SciPost Phys.\ Codeb.\ \textbf{2022} (2022), 8
doi:10.21468/SciPostPhysCodeb.8
\href{https://doi.org/10.21468/SciPostPhysCodeb.8}{[DOI]}




\bibitem{helenius2019epjc}
I.~Helenius and C.~O.~Rasmussen,
Eur.\ Phys.\ J.\ C \textbf{79} (2019) no.~5, 413
doi:10.1140/epjc/s10052-019-6914-1
\href{https://doi.org/10.1140/epjc/s10052-019-6914-1}{[DOI]}



\bibitem{Sjostrand:1987su}
T.~Sjostrand and M.~van Zijl,
Phys. Rev. D \textbf{36} (1987), 2019
doi:10.1103/PhysRevD.36.2019






\bibitem{paszke2019pytorch}
A.~Paszke, S.~Gross, F.~Massa, A.~Lerer, J.~Bradbury, G.~Chanan, T.~Killeen, Z.~Lin, 
N.~Gimelshein, L.~Antiga, A.~Desmaison, A.~Köpf, E.~Yang, Z.~DeVito, M.~Raison, 
A.~Tejani, S.~Chilamkurthy, B.~Steiner, L.~Fang, J.~Bai and S.~Chintala,
arXiv:1912.01703 [cs.LG]
\href{https://arxiv.org/abs/1912.01703}{[arXiv]}




\bibitem{Patro:2015Normalization}
S.~G.~K.~Patro and K.~K.~Sahu,
arXiv:1503.06462 [cs.OH]
\href{https://arxiv.org/abs/1503.06462}{[arXiv]}




\bibitem{Dubey:2021activation}
S.~R.~Dubey, S.~K.~Singh and B.~B.~Chaudhuri,
[arXiv:2109.14545 [cs.LG]].




\bibitem{Xu:2021vpn}
H.~j.~Xu, H.~Li, X.~Wang, C.~Shen and F.~Wang,
Phys. Lett. B \textbf{819} (2021), 136453
doi:10.1016/j.physletb.2021.136453
[arXiv:2103.05595 [nucl-th]].




\bibitem{Botchkarev:2018}
A.~Botchkarev,
doi:10.48550/arXiv.1809.03006










\end{thebibliography}
\end{document}